\documentclass[sigconf]{acmart}
\AtBeginDocument{%
  }

\copyrightyear{2026}
\acmYear{2026}
\setcopyright{cc}
\setcctype{by}
\acmConference[KDD '26]{Proceedings of the 32nd ACM SIGKDD Conference on Knowledge Discovery and Data Mining V.2}{August 09--13, 2026}{Jeju Island, Republic of Korea}
\acmBooktitle{Proceedings of the 32nd ACM SIGKDD Conference on Knowledge Discovery and Data Mining V.2 (KDD '26), August 09--13, 2026, Jeju Island, Republic of Korea}
\acmDOI{10.1145/3770855.3817447}
\acmISBN{979-8-4007-2259-2/2026/08}

\usepackage[utf8]{inputenc} %
\usepackage[T1]{fontenc}    %
\usepackage{hyperref}       %
\usepackage{url}            %
\usepackage{booktabs}       %
\usepackage{amsfonts}       %
\usepackage{nicefrac}       %
\usepackage{microtype}      %
\usepackage{xcolor}         %
\usepackage{xspace}
\usepackage{graphicx}
\usepackage{float}
\usepackage{subcaption}
\usepackage[most]{tcolorbox}
\usepackage{listings}
\usepackage{multirow}
\usepackage{pgffor}
\usepackage{xparse}
\usepackage{array}
\usepackage{wrapfig}
\usepackage{cleveref}
\usepackage{algorithm}
\usepackage{algpseudocode}

\tcbuselibrary{listingsutf8}

\newcommand{\pname}{\texttt{RELIANCE}\xspace}
\newcommand{\fullpname}{\texttt{Reproductive hEaLth Information Access in oNline Content Environments}\xspace}
\newcommand{\dataset}{\href{https://realize-lab.github.io/RELIANCE}{https://realize-lab.github.io/RELIANCE}}

\usepackage[table,xcdraw]{xcolor}

\newcommand{\gptfour}{\texttt{gpt-4o}\xspace}
\newcommand{\gptfourmini}{\texttt{gpt-4o-mini}\xspace}
\newcommand{\medgemma}{\texttt{medgemma-27B}\xspace}

\newcommand{\gemini}{\texttt{Gemini-2.0-Flash}\xspace}
\newcommand{\geminitwofive}{\texttt{Gemini-2.5-Flash}\xspace}
\newcommand{\qwen}{\texttt{Qwen}\xspace}

\newcommand{\qwenfourteen}{\texttt{Qwen3-14B}\xspace}
\newcommand{\openaiomni}{\texttt{omni-moderation}\xspace}
\newcommand{\llamaguard}{\texttt{Meta-Llama-Guard-3-8B}\xspace}
\newcommand{\shieldgemma}{\texttt{ShieldGemma-9B}\xspace}

\newcommand{\numqueries}{56\xspace}
\newcommand{\numvideos}{336\xspace}
\newcommand{\numsentences}{409\xspace}

\newcommand{\isaccurate}{\texttt{IsAccurate}\xspace}
\newcommand{\isharmful}{\texttt{IsHarmful}\xspace}

\newcommand{\greenbox}[1]{\colorbox[HTML]{d2e7d6}{#1}}

\newcommand{\purplebox}[1]{\colorbox[HTML]{D3D3FF}{#1}} %

\newcommand{\redbox}[1]{\colorbox[HTML]{FAA0A0}{#1}}
\newcommand{\lightbluebox}[1]{\colorbox[HTML]{609EE6}{#1}}

\newcommand{\q}[1]{``#1''}
\newtcolorbox{biggreenbox}[1][]{%
  colback=green!10!white,
  colframe=green!60!black,
  title=#1,
  sharp corners=south,
  breakable,
  enhanced,
  boxrule=0.8pt,
  left=5pt,
  right=5pt,
  top=5pt,
  bottom=5pt,
}
\newcommand{\inlineredbox}[1]{%
  \fcolorbox{red}{white}{\strut #1}%
}

\usepackage{fontawesome5} %

\newcommand{\searchtext}[1]{%
  \raisebox{0.5pt}{%
    \fcolorbox{gray!60}{gray!10}{%
      \scriptsize\sffamily\faSearch~#1%
    }%
  }%
}

\title{\pname: Curating and Evaluating Reproductive Health Information on Social Media}

\begin{document}

\author{Vaibhav Balloli}
\affiliation{%
  \institution{University of Michigan}
  \city{Ånn Arbor}
  \country{USA}}
\email{vballoli@umich.edu}

\author{Laura Peyton Ellis}
\affiliation{%
  \institution{University of Connecticut Health}
  \city{Farmington}
  \country{USA}}
\email{lauellis@uchc.edu}

\author{Vishala Mishra}
\affiliation{%
  \institution{Duke University}
  \city{Durham}
  \country{USA}}
\email{vishala.mishra@duke.edu}

\author{Alice Chi}
\affiliation{%
  \institution{University of Michigan}
  \city{Ånn Arbor}
  \country{USA}}
\email{amchi@med.umich.edu}

\author{Alex Peahl}
\affiliation{%
  \institution{University of Michigan}
  \city{Ånn Arbor}
  \country{USA}}
\email{alexfrie@med.umich.edu}

\author{Elizabeth Bondi-Kelly}
\affiliation{%
  \institution{University of Michigan}
  \city{Ånn Arbor}
  \country{USA}}
\email{ecbk@umich.edu}

\renewcommand{\shortauthors}{Vaibhav Balloli et al.}

\begin{abstract}
  A clear and well-documented \LaTeX\ document is presented as an
  article formatted for publication by ACM in a conference proceedings
  or journal publication. Based on the ``acmart'' document class, this
  article presents and explains many of the common variations, as well
  as many of the formatting elements an author may use in the
  preparation of the documentation of their work.
\end{abstract}



\begin{teaserfigure}
    \centering
    \includegraphics[width=0.6\linewidth]{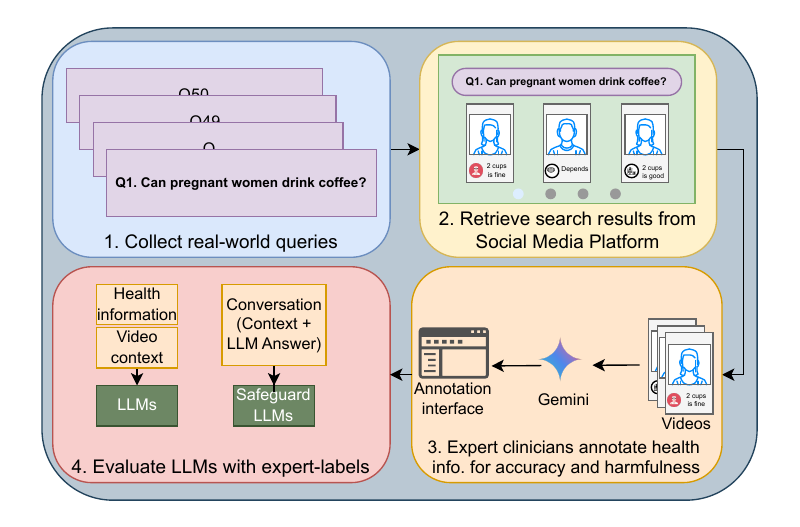}
    \caption{Overview of \pname. We collect a dataset of health information from TikTok to support real-world queries faced during pregnancy. We analyze this data and evaluate LLMs' capabilities in detecting inaccurate and harmful information.}
    \label{fig:data_collection}
    \Description{Illustration of our four-stage data collection process. 1) Collecting real-world queries pregnant people have during their pregnancy, 2) Collecting videos from TikTok by pasting these queries into the search bar, 3) Using Gemini-2.0-Flash to transcribe the audio, video, and text in the video and showing it to experts for annotation, and 4) General-purpose and Safeguard LLMs are evaluated on this data.}
\end{teaserfigure}

\begin{abstract}
    Social media platforms like TikTok have become a key source of health information, with studies reporting inaccuracies in posts. As Large Language Model (LLM) providers increasingly integrate LLMs into digital platforms to fact-check content (e.g., Grok and Perplexity on X and WhatsApp, respectively) and are being used by people to fact-check information, deploying these systems in critical areas such as reproductive health without rigorous evaluation can cause serious harm. We introduce RELIANCE, an expert-annotated dataset of health information on TikTok surrounding pregnancy and postpartum queries, serving as both an analysis of the reproductive health information landscape and an evaluation of LLMs' capabilities in fact-checking this content. Our dataset comprises 409 annotated sentences from 336 videos across 56 clinician-reviewed queries, annotated by three expert clinicians in Obstetrics, Gynecology, and Internal Medicine. Our findings reveal that nearly 60\% of the health information in the videos we sampled is accurate. Furthermore, LLM evaluations reveal a gap between evaluating specific claims and evaluating the entire content (15\%). We believe that our methodology, dataset, and tool will support the machine learning community in improving LLMs for important domains with real-world data, extending to other platforms and languages, and helping the health community further understand the information landscape on social media. Our dataset and code are made available \dataset.
\end{abstract}

\maketitle

\section{Introduction}
\label{sec:intro}

Digital platforms like social media have become ubiquitous in society, with 72\% of Americans surveyed using these sites for various purposes, including seeking health information \citep{pew2021}. For instance, people have used digital platforms to understand diagnoses \citep{zuccon2015diagnose}, listen to others' health experiences \citep{eysenbach2004health,balloli2026maternal}, and seek health advice \citep{johnson2006neo}. \citet{de2014seeking} finds that those accessing these digital platforms for health information cite  reasons like \q{\textit{convenience,}} \q{\textit{more detailed information 
or were dissatisfied with what a health care professional 
had previously told them,}} \q{\textit{sought other people’s recommendations, advice, or opinions.}} Similarly, deployed applications like \textit{ASHABot} \citep{ramjee2024ashabot}, a chatbot to support community health workers in pregnancy and reproductive health, found via qualitative analysis that at least one user turned to social media when they were unable to find relevant information through the bot. These platforms have also been linked to playing a key role in public health %
\citep{kanchan2023social}.

While there are potential benefits of connecting with others, the presence of and harms due to the inaccuracies on these platforms have been well documented in the literature \citep{wu2019misinformation,suarez2021prevalence,budak2024misunderstanding}. Inaccurate information on these platforms in critical areas such as pregnancy and reproductive health can have especially severe adverse consequences, creating \q{the perfect storm} due to vulnerabilities and historical biases \citep{antoniak2024nlp,balloli2026maternal}. In fact, leading organizations like the Centers for Disease Control (CDC) and American College of Obstetricians and Gynecologists (ACOG) have identified accurate and timely information as a key factor in the management of pregnancy and pregnancy-related complications \citep{cdcprevention,acogredesign}. Hence, understanding the accuracy and harmfulness of the reproductive health information available on these social media platforms can have a real-world impact on public health \citep{dunn2018social}.

Large Language Models (LLMs) have shown potential in mitigating misinformation, for example through fact-checking~\citep{tian2024web} and providing accurate information by integrating into chat applications~\citep{ramjee2024ashabot}. In the field of health, LLMs have shown great performance ($\sim$90\% accurate) in their question-answering (QA) and reasoning capabilities \citep{jin2021disease,hendrycks2020measuring,healthbench2025singhal}, assisting in reading clinical notes \citep{mannhardt2024impact}, medical reasoning and diagnoses \citep{saab2024capabilities}, and misinformation detection~\citep{sun2023med}. 

Yet, current health datasets and benchmarks for LLMs have largely focused on evaluating models on standardized medical exams or clinical questions and lack focus on the data available in the real world (like social media) and reproductive health in general. This is critically important to address, as LLM providers like Perplexity and xAI (Grok) are being integrated into major digital platforms like WhatsApp and X (formerly Twitter), and are actively prompting users to fact-check any information they receive using their LLMs (see Appendix \ref{sec:appendix_llm_platforms} for promotions on the use of LLMs in WhatsApp, YouTube, and Facebook for queries related to health and multimedia content). Parallel works like \cite{renault_mosleh_rand_2026,montero2026kff} report how users use these LLMs on X and in general to fact-check (\textbf{\textit{\q{$@$grok, fact-check this}}}) across various domains, including health and politics. In light of these moves, recent updates to major platforms like Meta have been scaling back human fact-checking programs, making LLM evaluations on social media content more important than ever \cite{kaplan2025meta}.

To support the machine learning community in improving LLMs for important domains with real-world data and helping the health community further understand the information landscape on social media, we present \pname (\fullpname): an expert-annotated dataset and evaluation framework for reproductive health information on TikTok. Our contributions are as follows: 

\begin{enumerate}
    \item A clinician-annotated dataset of reproductive health claims on TikTok. We collect the top six video search results for each of 56 clinician-reviewed queries on the peripartum period, yielding 336 videos. Four expert clinicians in Obstetrics, Gynecology, and Internal Medicine annotate \numsentences health-relevant claims for accuracy and harmfulness, capturing the important distinction that not all inaccurate information is harmful (\Cref{sec:dataset}).
    
    \item An analysis of the reproductive health information landscape on TikTok search. We find that peripartum content is widely viewed (averaging 273K views per video), approximately 30\% of the health claims surfaced by search are inaccurate, and roughly half of those inaccurate claims are clinically harmful. Engagement metrics (views, likes, followers) show no strong correlation with accuracy, suggesting popular content is no more reliable than less popular content (\Cref{sec:characteristics}).

    \item A granularity-sensitive evaluation of LLM fact-checking capabilities. We benchmark four LLMs on two tasks: claim-level verification and video-level inaccuracy detection. All models exhibit a consistent performance gap of approximately 15\% between these tasks, with \geminitwofive retaining the most performance at scale.

    \item An evaluation of safeguard models on reproductive health content. We assess three safety models and find that reproductive health queries are inconsistently flagged. For example, one model categorized \q{How much kicking should a pregnant woman feel?} as violence.
\end{enumerate}

\section{Background and Related Work}
\label{sec:background}

We will now provide background information and related work, starting with definitions of key terms we use throughout the paper, followed by additional background and related work on social media and natural language processing in reproductive health.

\begin{biggreenbox}[\large{\textbf{Definitions}}]
    \paragraph{\textbf{Reproductive health:}} We collectively refer to sexual and obstetric health as reproductive health. Example topics under reproductive health include conception, pregnancy, etc. \\
    \paragraph{\textbf{Peripartum period:}} Period before, during, and after pregnancy (postpartum) \citep{khan2023pregnancy}. \\
    \paragraph{\textbf{Inaccurate information:}} We define any health-related information not supported by scientific evidence or not part of standard clinical practice as \textit{inaccurate} information. \\
    \paragraph{\textbf{Harmful information:}}Any medically dangerous health information that causes adverse consequences.
\end{biggreenbox}

\begin{figure}
\centering
\begin{minipage}{0.49\linewidth}
    \begin{tcolorbox}[colback=brown!10!white, colframe=brown!80!black, title=Accurate: \faTimes \newline Harmful: \faTimes]
    \q{You should eat vegetables to make your baby smarter.}
    \end{tcolorbox}
\end{minipage}
\hfill
\begin{minipage}{0.49\linewidth}
    \begin{tcolorbox}[colback=red!10!white, colframe=red!70!black, title=Accurate: \faTimes \newline Harmful: \faCheck]
    \q{... our fertility cocktail will improve your fertility.}
    \end{tcolorbox}
\end{minipage}
\caption{Illustration of how inaccurate information can be harmless (left) or harmful (right).}
\label{fig:harmless_harmful}
\Description{Illustration of inaccurate information which is a) not harmful (e.g., You should eat vegetables to make your baby smarter) and b) harmful (e.g., our fertility cocktail will improve your fertility). }
\end{figure}

\begin{figure*}[t]
    \centering
    \begin{subfigure}[t]{0.6\textwidth}
        \centering
        \includegraphics[width=\linewidth]{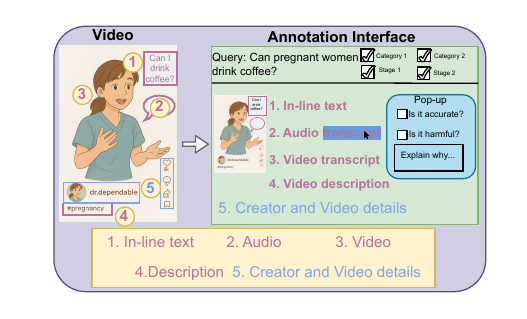} %
        \caption{Illustration of information extracted from a video and the corresponding expert annotation.}
        \label{fig:annotation}
    \end{subfigure}
    \hfill
    \begin{subfigure}[t]{0.35\textwidth}
        \centering
        \begin{tcolorbox}[colback=red!5!white, colframe=blue!75!black, title=Data row, width=\linewidth, boxsep=0pt, left=2pt, right=2pt, top=2pt, bottom=2pt]
\begin{lstlisting}[
  basicstyle=\small\ttfamily,
  breaklines=true,
  columns=fullflexible,
  escapeinside={|}{|},
  keepspaces=true,
  showstringspaces=false,
  tabsize=2
]
{
  "sentence": |\lightbluebox{'transcript'}|,
  "query": "Can pregnant...",
  "is_harmful": "No",
  "is_inaccurate": "No",
  "reason": "",
  "category" "1,2",
  "stage": "1,2"
}
\end{lstlisting}
\end{tcolorbox}
        \caption{Attributes from a single expert annotation}
        \label{fig:datarow}
    \end{subfigure}
    \caption{Annotation process and example illustration of a row in our \pname dataset.}
    \label{fig:annotation_with_datarow}
    \Description{On the left, the figure depicts how the TikTok video is processed into the annotation interface, where the interface has four attributes from the video: a) text presented in the video, b) audio transcript of the video, c) video transcript of the video, and d) video description posted by the creator. The experts use their mouse to select a sentence and annotate for accuracy and harmfulness, which category and stage the query belongs to, and any reasons explaining the inaccuracy or harm. These annotations are then exported into a file on the right-hand side of the image.}
\end{figure*}

\subsection{Social Media and Reproductive Health} We focus specifically on social media because  it has been a critical focus in understanding health information \cite{de2014seeking,gottfried2016news} and it frequently arises in the clinical experience of our co-author team. In the context of the United States, TikTok is one of the most accessed social media platforms, with reports of nearly 59\% of American adults under 30 using the platform \citep{pewtiktok}. Previous works investigate the role of various social media platforms like TikTok, YouTube, Facebook, etc., in reproductive health, including how healthcare professionals engage in public health discussions \citep{antheunis2013patients}, how care-seekers use social media as a source of information \citep{chen2021social}, how OB/GYNs (Obstetricians/Gynecologists) communicate
\citep{stein_examining_2022}, and what the content in a \textit{post-Roe} era contains \citep{smith_my_2024}. Collectively, these research findings show that \textbf{(a)} people consistently use social media to access health information, \textbf{(b)} social media is rife with misinformation, and \textbf{(c)} analyzing social media content and designing interventions to counteract it is important. We aim to build upon this work and develop new methods to continue these analyses more easily over time.

We also note that previous works have researched content on TikTok using {general \texttt{\#hashtags}} to retrieve the videos. For example, \citet{stein_examining_2022} uses \texttt{\#doctor,\#tiktokdoc, etc.} by manually assimilating popular hashtags used by doctors. We adopt a different lens for accessing the information on TikTok by utilizing the \searchtext{search bar} to search for information referencing specific real-world queries of care-seekers. We show later in the paper how this process can help analyze more specific content and complement these works by using topic-specific hashtags.

\subsection{Language Models and Misinformation} Natural Language Processing (NLP) and language model-based methods have the potential to improve question-answering capabilities around maternal health \cite{srikanth2023pregnant,nguyen2024rosie},  and have been successfully applied to detect misinformation \citep{perez2017automatic,chen2023can}, as well as reduce AI-generated misinformation \citep{penny2025reducing}. These works have demonstrated the capabilities of language models in designing interventions to reduce misinformation and promote safe information, which we aim to contribute to as well through this dataset. 

Our work is orthogonal and complementary to prior work in evaluating LLMs for healthcare tasks like MMLU \cite{hendrycks2020measuring} and MedQA \cite{jin2021disease}, which have considered question-answer-based evaluations on standardized tests and narrow clinical settings, as well as recent works like Healthbench \cite{healthbench2025singhal} and Med-MMHL \citep{sun2023med}, which have focused on textual data like human-LLM conversations and Twitter data, respectively. We aim to evaluate LLMs in identifying the accuracy and harmfulness of health information in real-world multimodal data on TikTok. By focusing on reproductive health information, we aim to contribute to improving their capabilities surrounding a historically understudied area of health \cite{national2024overview}.

\section{\pname Dataset}
\label{sec:dataset}

In this section, we provide details on data collection, annotation, and findings about our data. We refer the reader to the Appendix \ref{sec:appendix_annotation_design} for further details regarding techniques that did not work and additional reasoning behind specific choices on the platform and  LLMs.

\subsection{Data Collection}

\paragraph{Team:} Our team consists of four expert clinicians with experience in Obstetrics, Gynecology, and Internal Medicine who reviewed the queries used to curate our dataset. Following similar practices to the re-labeling of MedQA\citep{saab2024capabilities}, we collected three annotations per video. 

\paragraph{Queries:} In this dataset, we chose natural language questions similar to \citet{srikanth2023pregnant}. Previous works have shown that common $n$-grams ($n=2$ or $n=3$) that occur when searching the web are natural questions like \textit{How to} or \textit{Can I} \cite{volske_what_2015}. Additionally, \cite{srikanth2023pregnant} contains questions that were asked in the real world on Reddit\footnote{\url{reddit.com}} or information-seeking apps like Rosie \cite{nguyen2024rosie}. Since our focus was largely on peripartum health as opposed to also including infant health, we sourced questions from a mix of queries from \citep{srikanth2023pregnant}, ACOG, experience of the clinicians, and clinician-verified synthetically generated queries from \gptfour. Our clinical team selected the queries around the peripartum period that are the most realistic and likely to encounter in their clinical practice.

\paragraph{Platform data:} We collect the top six video search results for each of the \numqueries queries on TikTok\footnote{\url{tiktok.com}}. While all the data is collected from TikTok, we note that this procedure is extendable to other platforms through similar methods and accessible Application Programming Interfaces (APIs). We refer the reader to Appendix \ref{sec:platform_data} for more details on the restrictions and challenges faced in data collection.

\subsection{Annotation}

\paragraph{Transcription and Video attributes:} We use \gemini to transcribe the content of these videos. These transcripts are then transferred to the annotation interface for our expert clinical team to annotate. The interface contains the following information extracted from each video:

\begin{itemize}
    \item \textbf{Transcripts:} \gemini is used to generate three kinds of transcripts: (1) Audio transcript containing all of the information conveyed in the audio, (2) Video transcript containing all of the information conveyed through animations and visual effects, and (3) In-line text transcript for all of the text displayed in the video.
    \item \textbf{Video data:} In addition to the transcripts, we also extract the creator-led description for the video and metadata surrounding the video (views, likes, comments, and shares) and the creator (followers and likes across all creator videos).
\end{itemize}

\paragraph{Annotation attributes:} Figure \ref{fig:annotation} depicts the information extracted from a video and shown to our team. The annotation aims to tag sentences with medically relevant information, inaccurate information, and harmfulness. The hierarchy of annotation is as follows:

\begin{figure*}[t]
 \begin{subfigure}[t]{0.33\textwidth}
  \centering
  \includegraphics[width=\linewidth]{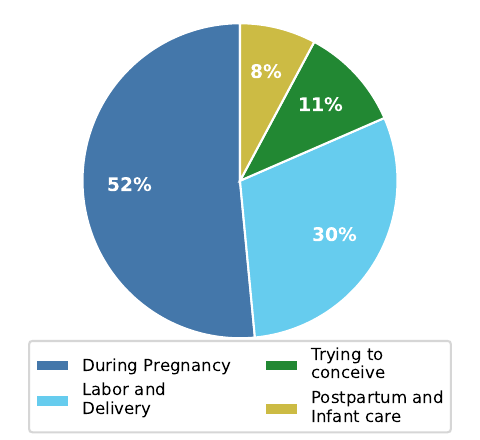}
  \caption{Stages Distribution}
  \label{fig:stage_distribution}
\end{subfigure}
\hfill
\begin{subfigure}[t]{0.33\textwidth}
  \centering
  \includegraphics[width=\linewidth]{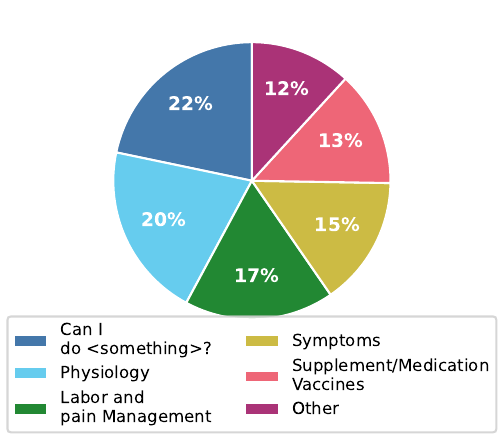}
  \caption{Categories Distribution}
  \label{fig:category_distribution}
\end{subfigure} 
\begin{subfigure}[t]{0.33\textwidth}
    \centering
    \includegraphics[width=\linewidth]{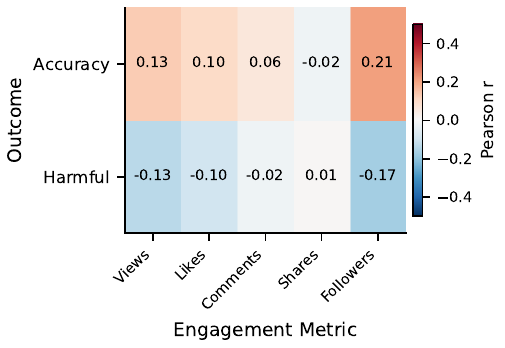}
    \caption{Engagement Analysis}
    \label{fig:engagement_correlation}
\end{subfigure}
\caption{Dataset analysis of distribution of annotations by (a) Stages, (b) Categories, and (c) correlation of labels with engagement metrics (likes, comments, shares, and creator followers).}
\label{fig:pname_all_in_one}
\Description{This figure has three subfigures. Figure A shows a pie chart showing the distribution of the dataset by the stages the queries belong to. Similarly, Figure B shows the distribution of the dataset based on the categories of the type of queries. Finally, Figure C shows the correlation of accuracy and harmfulness with engagement metrics of the video, like views, likes, shares, comments, and followers of the creator.}
\end{figure*}

\begin{itemize}
    \item Query-level annotation: Each query is tagged with the category and stage of pregnancy under which the query falls.
    \item Video-level annotation: For each of the six videos per query, \textit{medically-relevant} sentences are highlighted and tagged with \isaccurate and \isharmful
\end{itemize}

\begin{table}[t]
  \centering
  \begin{tabular}{cc}
\toprule
    Property & \%  \\ \midrule
    Accurate videos & 59.39 \\
    Harmful videos & 33.33 \\
    Accurate info & 69.91 \\
    Harmful info & 16.24 \\
    Search relevance & 88.9 \\
    Mean Annotator agreement (acc) & 82.97 \\
    Mean Annotator agreement (harm) & 86.91 \\
    Randolph's $\kappa$ (acc) & 75.22 \\
    Randolph's $\kappa$ (harm) & 80.76 \\
    \bottomrule
\end{tabular}

  \caption{Difference between aggregate label statistics, by full video and by info (per annotated information).}
  \label{tab:dataset_statistics}
\end{table}
\begin{table}[t]
    \centering
    \begin{tabular}{lccc}
\toprule
Property & Avg. & Median & Max. \\
\midrule
Views & 273K & 72K & 4M \\
Likes & 11K & 1K & 383K \\
Comments & 196 & 49 & 5K \\
Shares & 1K & 66 & 45K \\
Followers & 415K & 112K & 4M \\
Followers Likes & 12M & 2M & 241M \\
\bottomrule
\end{tabular}

    \caption{Statistics of the videos found across all queries.}
    \label{tab:video_stats}
\end{table}

After each expert selects a part of the transcript and annotates it with accuracy, harmfulness, and reason, it is saved as an individual data point in our \pname dataset (see Figure \ref{fig:datarow}). We aim to capture \isaccurate and \isharmful labels to capture the nuance that not all inaccurate information is harmful (Figure \ref{fig:harmless_harmful} illustrates the distinction). 
\paragraph{Processing and cleaning:} For each of the selected sentences/paragraphs, we first extract the complete sentence/paragraph from the original transcript and aggregate the labels. Furthermore, we de-identify the sentences to remove any personally identifiable information regarding the creators. Each of our experts spends 5-15 minutes carefully reviewing and annotating each query. Since each expert independently reviews the transcripts for sentences/paragraphs containing health information,  we add a processing stage to aggregate commonly labeled sentences. Algorithm \ref{algo:deanon} contains specific implementation details about data processing and cleaning. We only choose data that has two or more experts agreeing on the final label.

\subsection{Dataset Characteristics and Findings}
\label{sec:characteristics}

Our dataset \pname contains \numsentences sentences/paragraphs, each with a label for \isaccurate and \isharmful, from \numqueries queries and \numvideos videos. \Cref{tab:dataset_statistics,tab:video_stats} and Figure \ref{fig:pname_all_in_one} summarize the characteristics of our dataset.

\paragraph{Video metadata:} Table \ref{tab:video_stats} provides key statistics of interaction metrics of videos - \textit{views}, \textit{likes}, \textit{comments}, and \textit{shares}; and creators - \textit{followers} and \textit{net likes}. It is evident that \textbf{(a)} the content surrounding peripartum care has substantial visibility, averaging 273,000 views, and \textbf{(b)} the variance is high - with one creator having four million followers and the median creator having 112,000 followers. Figure \ref{fig:pname_all_in_one} depicts the distribution of the dataset in terms of stages of pregnancy, categories of queries, and correlation of metrics with respect to engagement metrics of the video.

\begin{table*}[t!]
  \centering
  \resizebox{\textwidth}{!}{%
  \begin{tabular}{@{}llrr@{}}
\toprule
\multicolumn{2}{c}{Queries with highest inaccuracies} &
  \multicolumn{2}{c}{Queries with highest harms} \\ \midrule
\begin{tabular}[c]{@{}c@{}}\purplebox{Can doctors force} \\ \purplebox{interventions during labor}?\end{tabular} &
  \begin{tabular}[c]{@{}c@{}}When does fundal height \\ correlate with gestational age?\end{tabular} &
  Do epidurals slow down labor? &
  \begin{tabular}[c]{@{}c@{}}\purplebox{Can doctors force} \\ \purplebox{interventions during labor?}\end{tabular} \\
\begin{tabular}[c]{@{}c@{}}When do many obstetricians\\ date the onset of pregnancy from?\end{tabular} &
  \begin{tabular}[c]{@{}c@{}}When should I take \\ folic acid in pregnancy?\end{tabular} &
  Can hot baths cause miscarriage? &
  \begin{tabular}[c]{@{}c@{}}\purplebox{How much kicking} \\ \purplebox{should pregnant women feel?}\end{tabular} \\
\begin{tabular}[c]{@{}c@{}}Is it safe to use essential \\ oils during pregnancy?\end{tabular} &
  \begin{tabular}[c]{@{}c@{}}\purplebox{When did they start testing} \\ \purplebox{for group b strep in pregnancy?}\end{tabular} &
  \begin{tabular}[c]{@{}c@{}}Are there other ways to \\ control pain besides epidural?\end{tabular} &
  \begin{tabular}[c]{@{}c@{}}\purplebox{When did they start testing} \\ \purplebox{for group b strep in pregnancy?}\end{tabular} \\
\begin{tabular}[c]{@{}c@{}}\purplebox{How much kicking} \\ \purplebox{should pregnant women feel?}\end{tabular} &
  \begin{tabular}[c]{@{}c@{}}Is it safe to travel by \\ plane during pregnancy?\end{tabular} &
  \begin{tabular}[c]{@{}c@{}}When do neural tube \\ defects occur during pregnancy?\end{tabular} &
  \begin{tabular}[c]{@{}c@{}}\purplebox{How can I improve} \\ \purplebox{my fertility naturally?}\end{tabular} \\
\begin{tabular}[c]{@{}c@{}}Is it normal to experience \\ Braxton hicks at 30 weeks?\end{tabular} &
  \begin{tabular}[c]{@{}c@{}}\purplebox{How can I improve} \\ \purplebox{my fertility naturally?}\end{tabular} &
  \begin{tabular}[c]{@{}c@{}}Is it safe to \\ have a baby at home?\end{tabular} &
  \begin{tabular}[c]{@{}c@{}}Can listening to music \\ make your baby smarter?\end{tabular} \\ \bottomrule
\end{tabular}%

  }
  \captionof{table}{Top 10 queries that resulted in the highest number of inaccurate and harmful information, respectively. \purplebox{Common queries between both of these labels are highlighted.}}
  \label{tab:top_queries}
\end{table*}

\paragraph{Information quality and relevance:} Table \ref{tab:dataset_statistics} shows key metrics about the quality of the health information. We find that approximately 40\% of the health information found through search is inaccurate, and about 75\% of it is harmful. Table \ref{tab:top_queries} lists the top ten queries with the highest inaccuracies and harmful information. We find that only four queries occur in both lists, highlighting the necessity to label inaccurate and harmful information separately. The data we analyze has high Randolph's $\kappa$ and mean annotator agreement. Note that our findings here approximately match with other works \cite{aaron2023labor} investigating health content also find similar levels of accuracy of information.

\begin{biggreenbox}[\textbf{Key Dataset Findings}]
\textbf{Search uncovers relevant videos:} Videos found via search are largely relevant and validate existing studies on the effectiveness of using search on social media to find information.\\

\textbf{Pregnancy and health content is popular:} Reproductive health information is popular (on the scale of 100K magnitude of views), suggesting the importance and need for monitoring critical information at scale.\\

\textbf{Misinformation is pervasive:} We find that the content we uncover using search has inaccurate and harmful information, indicating the need to classify and tag both labels and measure the intervention's capability of detection accordingly.\\

\textbf{Video popularity is not predictive of accuracy:} \textbf{There is no strong correlation} of accuracy or harm with metrics like likes, views, and followers, suggesting susceptibility of popular videos to having misinformation.

\end{biggreenbox}

\begin{figure*}[t]
  \centering

  \begin{subfigure}[t]{0.49\textwidth}
    \centering
    \includegraphics[width=\linewidth]{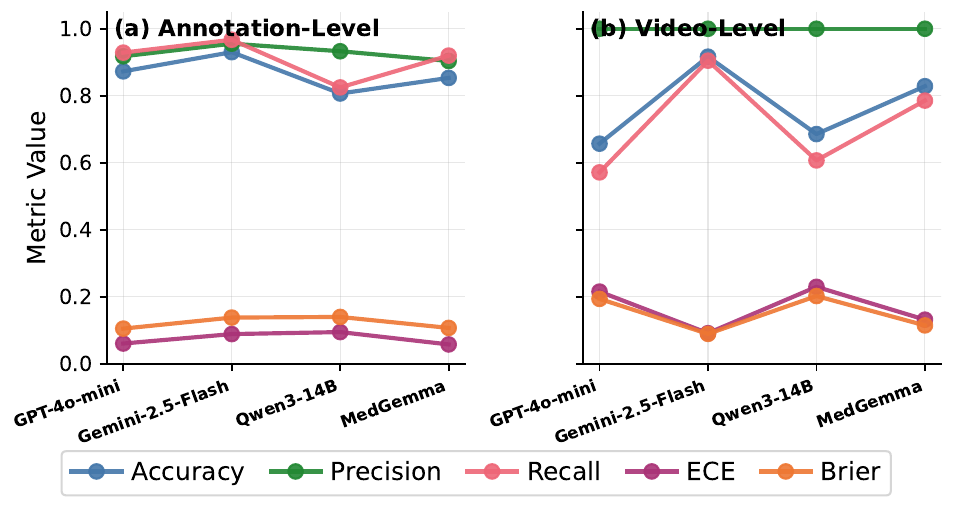}
    \caption{Performance of LLMs in detecting inaccurate information.}
    \Description{Two }
    \label{fig:llm_eval_inacc}
  \end{subfigure}\hfill
  \begin{subfigure}[t]{0.49\textwidth}
    \centering
    \includegraphics[width=\linewidth]{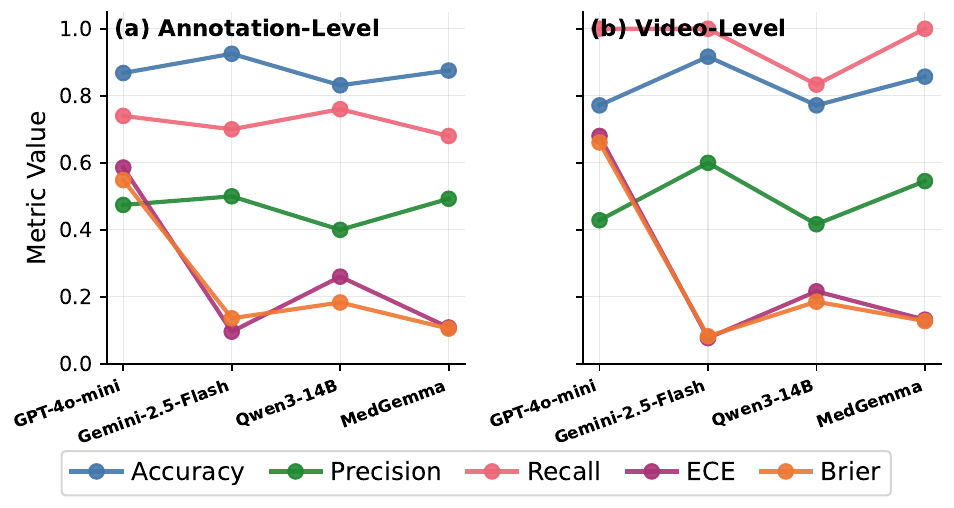}
    \caption{Performance of LLMs in detecting harmful information.}
    \label{fig:llm_eval_harm}
  \end{subfigure}

  \caption{Performance metrics (accuracy, precision, recall) and calibration metrics (ECE and Brier Score) on detecting inaccurate and harmful information.}
  \label{fig:llm_eval}
\end{figure*}

\section{LLM Evaluations}
\label{sec:evaluations}

This section investigates how LLMs detect the inaccuracy and harmfulness of the sentences/paragraphs in \pname. Understanding how LLMs perform on this data is crucial in interventions and applications that are designed for various stakeholders in the field of reproductive health (\cite{ramjee2024ashabot}). We consider two different types of LLMs:

\begin{itemize}
    \item \textbf{General-purpose LLMs (GP-LLMs):} We collectively refer to LLMs designed to support a large range of tasks and trained via pre-training and post-training/instruction tuning as general-purpose LLMs (GP-LLMs). These are also referred to as post-trained LLMs, that is, LLMs that have additional training to improve alignment and safety when used by users.
    
    \item \textbf{Safeguard LLMs:} General-purpose LLM responses have been generally shown to pose various risks like hallucinations \cite{mckenna2023sources} and harmful information \cite{liang2022holistic}. Safety researchers and LLM providers \citep{markov2023holistic} have called for robust moderation techniques for both user inputs and LLM-generated responses. Yet, harmful speech detection suffers from biased false positives \cite{dorn2024harmful}. Our aim in including these models is to understand the types of queries, content, and responses that are flagged, since these decisions during deployment have direct consequences for the developer and the user.

\end{itemize}

While both categories have numerous LLMs available, we chose a balance of closed-weight and open-weight models (that are capable of running on a single 80GB GPU) with a similar number of parameters to benchmark and analyze. All results have been reported by averaging three runs with separate seeds and temperature 0.5.

\subsection{Can LLMs Detect Inaccuracies and Harms at Various Granularities?}

We identify that misinformation detection can occur at two stages of granularity: \textbf{(Video-level):} The user can share the video directly to the LLM and ask if it contains any misinformation, and \textbf{(Claim-level):} The user can share the video and specifically ask for verification on a specific claim made in the video. These levels of granularity reflect various practical methodologies for user preferences in consuming information and verification through LLMs.

\paragraph{Performance.} Figure \ref{fig:llm_eval} compares LLMs for their performance and calibration scores on detecting inaccurate and harmful information. Precision here refers to the fraction that was truly accurate or harmful, and recall refers to the fraction that the model labeled as accurate or harmful. Expected Calibration Error (ECE) and Brier scores quantify the difference in predicted probabilities and ground truth, where the former is measured by binning, and the latter is measured by mean square error (the lower, the better). We first note that \geminitwofive performs the best, with high true positive and false negative scores. Surprisingly, the state-of-the-art medical model \medgemma performs similarly to the \qwen family of models and \gptfourmini. The table also highlights the ability of most LLMs to consistently perform better at detecting harm than inaccuracies. A key observation that stems from this is the gap in performance of LLMs in identifying inaccuracies and harms from a large context as opposed to verifying a particular claim, where a majority of the LLMs used have a huge drop in performance, with \geminitwofive retaining most of its performance. We hypothesize this is likely due to the good performance of models in the \texttt{Gemini} family, which have been shown to perform well in high token context windows.

\paragraph{Calibration.} In addition to prompting the LLMs to output accuracy and safety labels, the prompt also contains instructions to verbalize the confidence estimates of LLMs\cite{yang2024verbalized}, to evaluate how calibrated the models are with respect to their decisions on accuracy and harm. Note that these confidence estimates are elicited for both accuracy and harmful labels. We find that the models are well calibrated in general, aligning with observations previously made on LLMs, largely having the ability to know what they know \cite{kadavath2022language}. Interestingly, we note that \medgemma and \geminitwofive had the lowest ECE and Brier scores, indicating better calibration as opposed to other models.

\begin{figure}
    \includegraphics[width=0.75\linewidth]{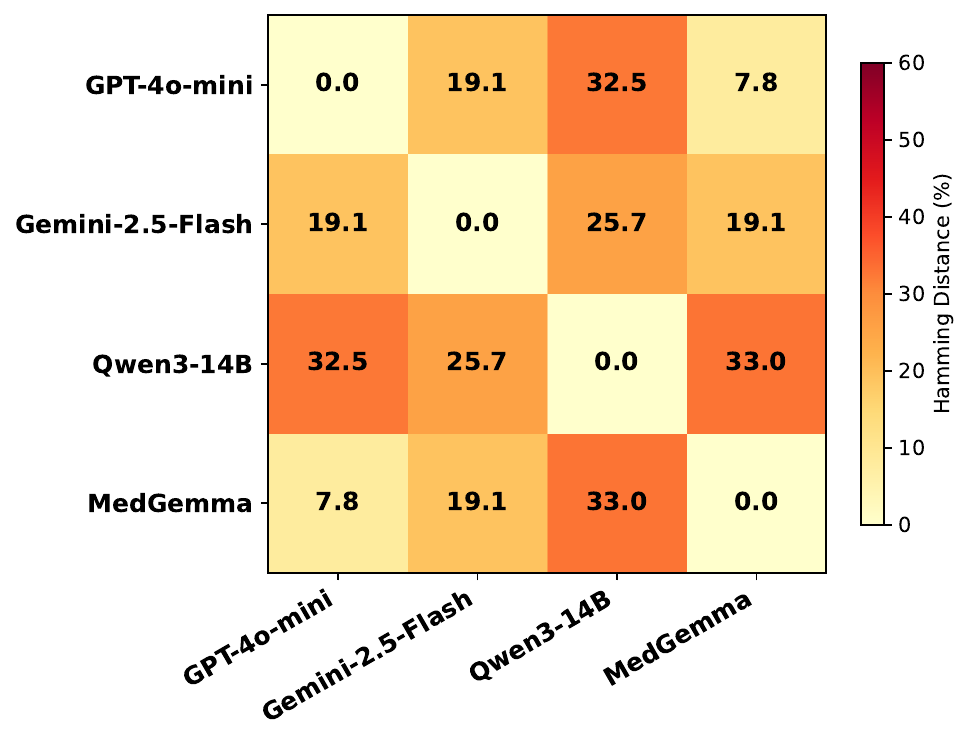}
    \caption{Hamming distance between LLM predictions.}
    \label{fig:hamming_accuracy}
\end{figure}

\paragraph{Multi-agent voting.} An obvious extension of using individual LLMs is to aggregate multiple LLMs to improve performance. Figure \ref{fig:hamming_accuracy} compares the Hamming distance (number of positions where two binary vectors differ) between all the LLMs for the \isaccurate task, illustrating how close or far apart these models are from each other (lower indicating closer). This illustrates that the capabilities of these models differ a lot from each other, suggesting that multiple expert LLMs cannot be chosen at random and require careful selection. Using a simple subset selection, we find that the best ensemble of three LLMs is \geminitwofive, \medgemma, and \qwenfourteen.

\begin{figure}
    \centering
    \includegraphics[width=\linewidth]{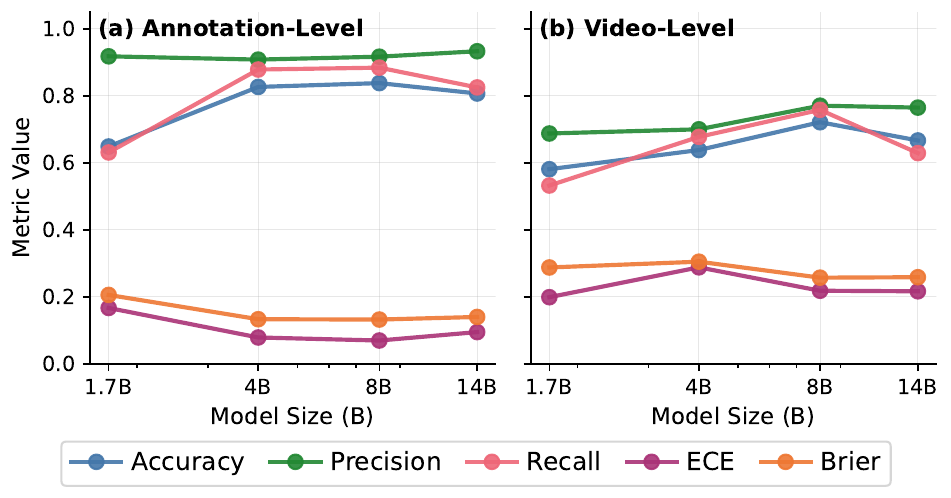}
    \caption{Comparing all metrics against the scale of models of the same model family (\qwen).}
    \label{fig:scaling}
\end{figure}

\paragraph{Scaling.} We evaluate the \qwen family of models (1.7B, 4B, 8B, and 14B), state-of-the-art open-source models on our dataset, to understand if the performance metrics we have evaluated until now change with the scale of the model, since these models have been trained using similar techniques on the same amount of data \cite{yang2025qwen3}. Figure \ref{fig:scaling} tracks accuracy, precision, recall, and calibration scores across the four models. We notice that the trends are largely flat across both tasks across most metrics, suggesting that for critical domains like reproductive health, scaling the quality of available data used in pre-training and post-training of these models should help improve their performance.

\subsection{Can Safeguard Models Appropriately Handle Reproductive Health Content?}

We evaluate fine-tuned moderation models \openaiomni\citep{openaiomni} (API-based access), \llamaguard\citep{dubey2024llama3herdmodels}, and \newline \shieldgemma\citep{zeng2024shieldgemma} (open-weight access). Again, we chose these two open-access models since they have a similar number of parameters.
\begin{table}[t]
\centering
\begin{tabular*}{\columnwidth}{@{\extracolsep{\fill}}lcc@{}}
\toprule
\multirow{2}{*}{LLMs / \% False Positive} & \multicolumn{2}{c}{Safety task} \\
 & Flags user question & Flags LLM response \\
\midrule
\openaiomni  & \redbox{3.57} & \greenbox{0.1} \\
\llamaguard  & \greenbox{0.} & \redbox{4.2} \\
\shieldgemma & \greenbox{0.} & \greenbox{0.5} \\
\bottomrule
\end{tabular*}
\caption{Performance of Safeguard LLMs on flagging user questions and LLM responses for violating their internal guidelines.}
\label{tab:safeguard}
\end{table}

We consider two safety evaluations consistent with the principles of these models: \textbf{(Task-1) Flagging user question and video transcript}, that is, the user prompts the LLM to identify inaccuracies and harms in the video transcript surrounding their query and \textbf{(Task 2) Flagging LLM conversation}, that is, classifying the \textit{entire} context and LLM-response to the user query and transcript as a violation of safety guidelines. These models have been trained differently to account for different safety guidelines and inputs; therefore, the numbers presented in Table \ref{tab:safeguard} are standalone and not meant for comparison. 

Upon inspecting the categories for violation, each of the three flag different queries and contexts for various violations. For instance, \openaiomni flags queries like \textit{\q{How much kicking should a pregnant woman feel?}} under \textit{\q{violence}}. On the other hand, \llamaguard mostly flags LLM responses under \textit{specialized advice}, which we argue has a vague definition in this context since most LLM responses contain medical information that is relevant to answering the query, indicating most or all responses should be flagged under \textit{specialized advice}. This highlights the need for additional fine-tuning with this context in mind to ensure consistency in predictions. Finally, \medgemma performs the best with rare false positives.

\begin{biggreenbox}[\textbf{\large{Key LLM Evaluation Findings}}]

\textbf{Context window and task construction matters:} LLMs have a performance gap between evaluating specific statements and the entire context, suggesting multiple iterations of checks for each context during deployment.\\

\textbf{Current evals suggest multi-agent systems increase performance:} Multi-agent systems are largely beneficial, but require careful construction based on the pluralistic needs of the users during deployment.\\

\textbf{Scaling models is not the bottleneck:} Scaling helps only to a certain extent, but performance quickly plateaus, indicating the need to improve pre-training and post-training data in underrepresented domains and to handle queries accordingly.\\

\textbf{Safeguard models should be carefully chosen:} Different safeguard models provide different features (and lead to different outcomes), suggesting the need for comprehensive evaluations with such models for research and deployment with real users.
\end{biggreenbox}

\section{Discussion and Future Work}
\label{sec:discussion_future_work}

Our contributions critically analyze the information and LLM landscape and enable scaling data mining and analysis for future research and researchers. We now discuss the key learnings from our work and its implications for other domains and researchers in the real-world, including policy design, intervention, and deployment design, and acknowledge the limitations of this work.

\paragraph{Learnings:} \textbf{(a) Scalable methodology:} We show in Section \ref{sec:characteristics} that our methodology uncovers topic-specific relevant information by using search to find data, LLMs-in-the-loop to process the data, and experts to validate it, which acts as a starting point for future research extending to other domains. \textbf{(b) Large Language Models still have a long way to go:} All LLMs we tested demonstrate impressive performance on medical question-answering benchmarks like MedQA \cite{jin2021disease}, MMLU (Clinical Knowledge) \cite{hendrycks2020measuring}, and many more. However, our results demonstrate that quality data in the area of reproductive health is needed to improve the abilities of these models to steer users away from inaccuracies and harms. This raises questions about how publicly available LLMs handle domains that are underrepresented during training and are calibrated for deployment.

\paragraph{Policy Implications.} Our work has direct relevance to the evolving moderation landscape of content on platforms. Meta's January 2025 decision to end its third-party fact-checking program in the United States \cite{kaplan2025meta} shifts greater responsibility onto automated systems and community-based methods for content verification. Our evaluation shows that current LLMs are not yet reliable enough for autonomous deployment in this role, with the best-performing model (\geminitwofive) achieving only 84.2\% accuracy in identifying inaccuracies at the video level, and performance dropped significantly for other models. The 2025 federal restructuring that disrupted key CDC programs for maternal health surveillance \cite{kff2025disparities} underscores the need for complementary, community and technology-driven approaches to understanding the reproductive health information landscape, similar to Rosie\cite{nguyen2024rosie}.

\paragraph{Intervention Design.} Our findings highlight the need for principled system design in information-support applications that integrate multimodal sources, such as social media videos, to better serve users. We also believe our methodology can be extended to design community-sourced verification systems with LLMs in the loop, helping community stakeholders access factual and safe information and complementing existing work such as Rosie \cite{nguyen2024rosie}.

\paragraph{Limitations:}
\label{sec:limitations}Our contributions are subject to a few limitations. Even though we uncover highly relevant data, our findings extend to videos that actively appear when they are searched for. While recommendation algorithms play a key role in what content is shown, it is intractable to collect user data without data donation or direct access. Second, we define our scope to English content within the United States on TikTok due to our annotator expertise and budget. We acknowledge that different countries have different platforms that are popular, and these findings might not transfer directly. Hence, we release our tool and describe our methodology extensively to enable extending this to other languages and platforms. Finally, this analysis is only a snapshot of the search at the time of data collection (January 2025). Social media platforms are dynamic and ever-changing, and we hope that this provides a starting point for long-term studies.

\section*{Ethics Statement}
\paragraph{Institutional Review and Anonymity.} Our research proposal was reviewed by the Institutional Review Board at the authors' institution and was declared \textit{exempt} from oversight. Although our work was exempt, we acknowledge the concerns of using \textit{publicly available data} \cite{buck2021didn,zimmer2020but,stephenson2024sharenting}. In light of these concerns and recommendations listed by \citet{antoniak2024nlp} calling for authors to clarify the implications of research works in NLP and reproductive health, we follow procedures (automated and manual) to de-identify all data we collect and emphasize that we intend to \textbf{(1)} understand the information landscape surrounding reproductive health, \textbf{(2)} evaluate the capabilities of current LLMs in detecting inaccuracies and \textbf{(3)} provide methodologies for future research. 
\paragraph{Intent.} We recognize that social media content often contains information and inferences from personal anecdotes and experiences. Our clinicians carefully annotated based on their medical judgment and experience, and have no intention to invalidate any of the creators' experiences. We also re-emphasize that this work is not a reflection of the platform or the content creators, but rather a scientific analysis of characteristics of reproductive health information in social media at large, and how LLMs perform. 

\section*{Acknowledgments}
This work was partially supported by funding from Google and the University of Michigan (including the Raoul Wallenberg Institute, E-Health and Artificial Intelligence, and the Center for Academic Innovation). Alex Peahl was also supported by Pulsenmore, NICHD, BCBS of Michigan, MHEF, and Molina Foundation. Vishala Mishra would also like to thank Vasudha Mishra for helpful discussions and feedback.

\bibliographystyle{abbrvnat}
\bibliography{references}

\appendix

\section{Design choices and artifacts}
\label{sec:appendix_design_choicces_artifacts}

\subsection{Release Artifacts}
\label{sec:appendix_artifacts}

We release the data on \dataset under the \texttt{CC-BY-NC 4.0} license.

    \begin{itemize}
        \item Sentences/paragraphs annotated by our experts.
        \item Aggregated \isaccurate and \isharmful labels for each of the sentences/paragraphs.
        \item Categories and stages associated with the query.
        \item Code to collect similar data.
    \end{itemize}

\subsection{Annotation design}
\label{sec:appendix_annotation_design}

\paragraph{Moving away from FactScore-style of evaluation:} FactScore \cite{min2023factscore} is an evaluation strategy designed for long-form text. This method breaks paragraphs into a series of atomic facts and computes the percentage of atomic facts that are supported by a grounded knowledge source. Our initial annotation design included using LLMs to break the transcripts into atomic facts and annotating them for accuracy and harms. However, we found that the atomic facts generated had the following issues:

\begin{itemize}
    \item \textit{Inconsistent phrasing:} The atomic facts that were generated often were inconsistent and sometimes \textit{wrong}
    \item \textit{Missing context:} Atomic facts generated often had the larger context associated with the fact from a previous sentence missing.
\end{itemize}

Furthermore, we were unable to fix this issue with different LLMs and few-shot prompting techniques. Our hypothesis as to why this happened is the nature and content of text, the LLMs were not able to perform as well as we wanted them to, and thus we reverted to our manually selecting the text.

\subsection{Additional Data Collection Details}
\label{sec:platform_data}

\begin{tcolorbox}[colback=red!5!white,colframe=red!75!black,title=TRIGGER WARNING]
This contains some text related to self-harm. Readers are advised to proceed with care if this distresses them.
\end{tcolorbox}

\paragraph{Choice of TikTok:} TikTok is emerging as a popular platform to access content in the United States. Multiple research works have suggested how experts amass a large following on social media platforms like TikTok\citep{kwon2023obstetric}, with this space often referred to as \texttt{MedTok} \cite{rao2022medfluencing}). Hence, we chose TikTok to better understand accessing specific information on the questions surrounding the peripartum period. Furthermore, creators on social media platforms are not usually exclusive to a specific platform and often \textit{cross-post} the same content across multiple platforms \cite{farahbakhsh2015characterization}. This makes the information uncovered here likely to be found on other platforms popular in the creators' surroundings.

In order to validate these, we randomly sample 30 creators and their videos and manually find their content on other platforms. We were able to find 20 creators with the same usernames on other platforms like YouTube, Instagram, Facebook, and podcast platforms like Spotify and Apple Podcasts. Notably, Instagram and YouTube were the most frequent, with 60\% rate of having both a TikTok account and an Instagram account. We note that finding the same video on different platforms occurred 56.6\% of the time, indicating the preferences of content creators in posting the same or different content based on their target platform and audience.

\paragraph{Accessing data:}Following \cite{stein_examining_2022}, we create a fresh account on TikTok to access the search results with the least amount of bias. Search results for the same queries at different times returned almost similar results. 

\paragraph{Choice of \gemini:} We specifically choose this model for all the data processing requirements (transcript generation in Figure~\ref{fig:annotation}) due to the large context window support for multimedia content (videos). As of the date of video collection, no other multimodal models supported the length of these short-form videos. Furthermore, \gemini rate limits suited our data collection goals. We manually verified the accuracy of \gemini to reliably transcribe 20 random videos before using it on the entire dataset. 

\paragraph{Multi-modal capabilities:} We chose to transcribe to ensure the models are compared on the same capabilities across the same input data. However, we also find that prompting \gemini to first transcribe the content and then classify recovers similar performance.

\paragraph{Agreement:} When labels are aggregated for each video, the annotators had an average agreement of 67.62\% and 76.19\% for \isaccurate and \isharmful labels, respectively.

\begin{figure}
    \centering
    \includegraphics[width=0.7\linewidth]{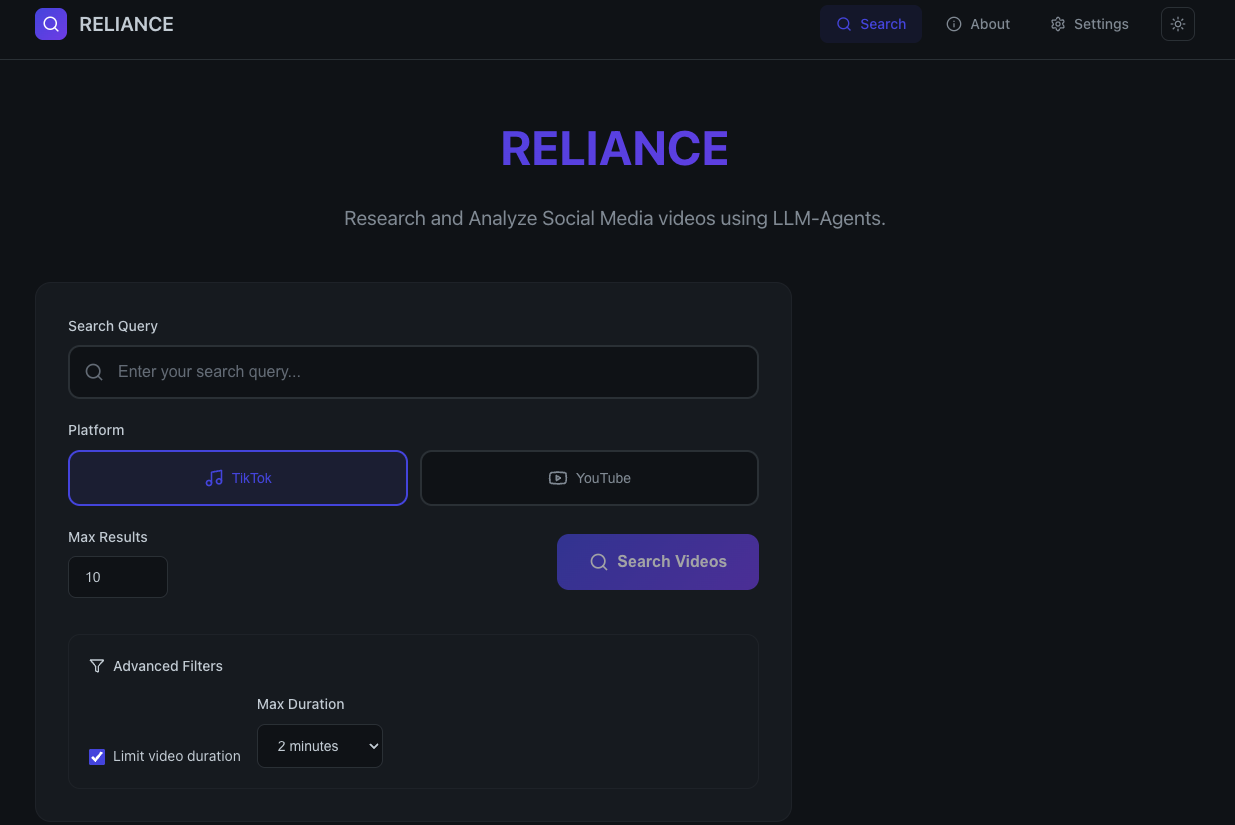}
    \caption{Screenshot of the application to collect data from TikTok.}
    \label{fig:app}
\end{figure}

\paragraph{Code to collect similar data:} We open-source the application and code to collect data per search query. Figure \ref{fig:app} depicts a screenshot of the application.

\begin{figure}
    \centering
    \begin{minipage}[b]{0.48\textwidth}
        \centering
        \begin{subfigure}[b]{\textwidth}
            \includegraphics[width=0.8\textwidth]{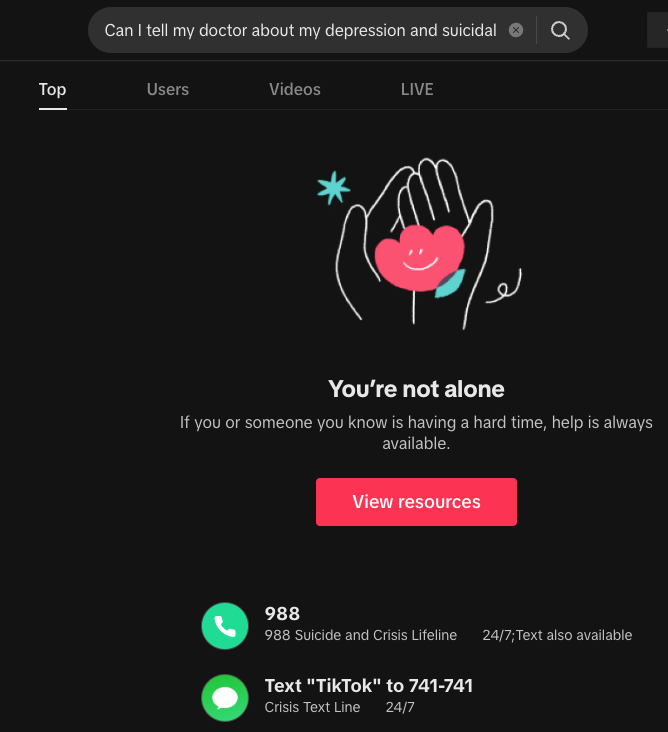}
            \caption{Warning when a query is related to self-harm.}
            \label{fig:tiktok_guideline_selfharm}
        \end{subfigure}
    \end{minipage}
    \hfill
    \begin{minipage}[b]{0.48\textwidth}
        \centering
        \begin{subfigure}[b]{\textwidth}
            \includegraphics[width=0.8\textwidth]{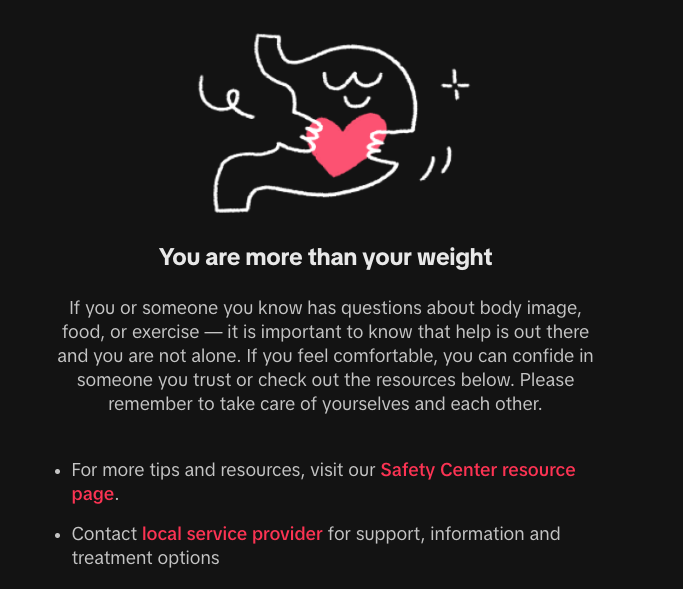}
            \caption{A message regarding weight-positivity for a query regarding weight gain during pregnancy.}
            \label{fig:tiktok_guideline_weight}
        \end{subfigure}
        
        \vspace{1em}
        
        \begin{subfigure}[b]{\textwidth}
            \includegraphics[width=0.8\textwidth]{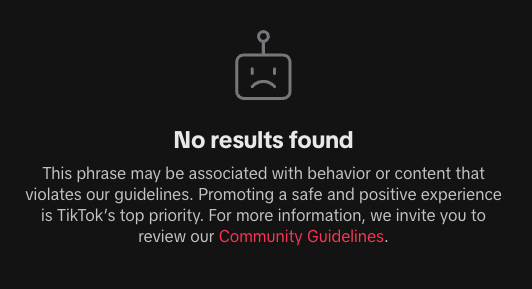}
            \caption{TikTok guidelines page for queries that violated platform policies, which occurred on two queries.}
            \label{fig:tiktok_guideline_comunity_guideline}
        \end{subfigure}
    \end{minipage}
    \caption{Combined figure showing Image 1 on the left, and Images 2 and 3 stacked on the right.}
    \label{fig:combined}
\end{figure}

\paragraph{Guardrails in TikTok:} Certain queries were marked as violating platform guidelines / standardized interventions to make sure the user is okay. Figure \ref{fig:combined} displays instances of such results.

\paragraph{Data aggregation and anonymization:} Algorithm \ref{algo:deanon} depicts our process of clustering, label aggregation, and removing PII data. Clustering is performed since different annotators can choose the same sentence piece in various ways.

\begin{algorithm}[H]
\caption{Information Label Aggregation and Deanonymization}
\label{algo:deanon}
\begin{algorithmic}[1]
\Require Selected information inputs $I$
\Ensure Deanonymized output labels

\State \textbf{FindParagraph}: Locate the paragraphs $P$ that contains the selected information inputs $I$
\State \textbf{GroupSimilar}: Group lexically similar paragraphs $P$
\State \textbf{AggregateLabels}: Collect and aggregate associated labels $\{L\}$ with the lexically similar paragraphs $P$
\State \textbf{MajorityVote}: Determine final label $L^*$ by applying a majority vote over $\{L\}$
\State \textbf{Anonymize}: Use \textsc{Presidio}\footnote{\url{https://microsoft.github.io/presidio/}} on $P$ with label $L^*$ to remove Personally Identifiable Information (PII)

\end{algorithmic}
\end{algorithm}

\subsection{Annotator Prompts}
\label{sec:appendix_annotator_prompts}

\begin{biggreenbox}[Instructions]
    The annotation interface contains \numqueries queries, each query with six videos (on the left of the screen) and corresponding audio, visual transcripts, and description on the right. Drag your mouse with a left click and release the button to select a medically relevant sentence/paragraph and annotate it for medical accuracy and harms (information that can cause adverse consequences). Please use your medical judgment to decide the accuracy and harms. Use the switch buttons to indicate if the entire content is irrelevant. 
\end{biggreenbox}

\subsection{Prompts and Evaluation Details}
\label{sec:appendix_prompts_evaluation}

\subsubsection{Transcription Prompts}
\begin{biggreenbox}[\gemini Audio Transcrbe Prompt]
\q{You're a helpful video processing assistant. Transcribe the audio in the given video and return the verbatim content; ignore any background music. Do not modify the content and give a verbatim audio transcription.}
\end{biggreenbox}

\begin{biggreenbox}[\gemini Video Transcribe Prompt]
    \q{You're a helpful video processing assistant. Give a verbose description of what the video is showing - describe the animations, if any, describe the actions the people in the video are doing or taking, describe the setting of the video. Give a verbose description in third-person perspective.}
\end{biggreenbox}

\begin{biggreenbox}[\gemini In-line text Transcribe Prompt]
    \q{You're a helpful video processing assistant. Transcribe any text that is displayed in the video and return the verbatim content, ignore the audio. Do not modify the text in any way and return as seen in the video.}
\end{biggreenbox}

\subsubsection{Evaluation Prompts}

\begin{biggreenbox}[LLM evaluating health content prompt]
    \q{Given a text and the full context, provide detailed reasoning on its accuracy and harmfulness and return the answer in the following format: ACCURATE: <yes/no> HARMFUL: <yes/no> REASON: <reason>. Context: {context} \textbackslash n\textbackslash n Text: {text} \textbackslash n\textbackslash n.}
\end{biggreenbox}

\subsubsection{Evaluation Hyperparameters}

Temperature is set to 0.5 with their respective maximum output length limit when generating responses for all LLMs (including LLMs for transcription and evaluation and the safeguard LLMs).

\subsubsection{Safeguard LLM Evaluation}

We use the procedures and prompts verbatim from model cards/descriptions of \openaiomni, \newline \llamaguard, and \shieldgemma provided by the developers. We also use the default guidelines that are included in the chat template with \llamaguard and \shieldgemma to make sure we only test for the guidelines these models were trained on.

\subsubsection{Compute}
\label{sec:compute}

All our LLM-related experiments were run on a workstation with three \texttt{NVidia RTX A6000 48GB GPUs} with 500GB RAM. Our annotation interface was hosted on another workstation with an \texttt{Intel i7} processor with 16GB RAM.

\section{Additional results}

\begin{figure}[H]
\begin{minipage}{\linewidth}
\centering
\begin{table}[H]
\centering
\resizebox{\textwidth}{!}{%
\begin{tabular}{@{}ccccc@{}}
\toprule
\multicolumn{1}{c}{\multirow{2}{*}{Task-Metric}} & \multicolumn{4}{c}{LLMs / \% Accuracy} \\ \cmidrule(l){2-5} 
\multicolumn{1}{c}{}                             & \gptfourmini   & \texttt{Gemini-2.5-Flash}   & \qwenfourteen   & \medgemma  \\ \midrule
Annotation-level (detect inaccuracy)              &  80.19  &  \textbf{87.32}  &  77.19  &  80.29  \\
Video-level (detect inaccuracy)              &  67.67  &  \textbf{84.21}  &  66.92  &  68.42  \\
Annotation-level (detect harm)              &  81.32  &  \textbf{89.10}  &  82.46  &  84.73  \\
Video-level (detect harm)              &  69.17  &  \textbf{85.71}  &  74.44  &  72.93  \\
\bottomrule
\end{tabular}%
}
\caption{LLM accuracy (\%) on identifying inaccuracy and harms across annotation-level (Task-1) and video-level (Task-2)}
\label{tab:classification}
\end{table}
\end{minipage}
\end{figure}

\begin{table}[H]
  \centering
  \begin{tabular}{lcc}
  \toprule
  Benchmark & $\Delta$ Accuracy & $\Delta$ Harmful \\
  \midrule
  Sentence-level & $0.0$ & $-3.2$ \\
  Video-level    & $-3.7$ & $-3.7$ \\
  \bottomrule
  \end{tabular}
  \label{tab:search-ablation-delta}
  \caption{Effect of enabling Google Search grounding on \texttt{gemini-2.5-flash} since \gemini was deprecated at the time of publication. $\Delta$ is search-on minus search-off, in percentage points on 30 samples. We hypothesize that searching for niche sentences can lead to the LLM hinging on untrustworthy webpages.}
\end{table}

\begin{figure*}
    \includegraphics[width=0.8\linewidth]{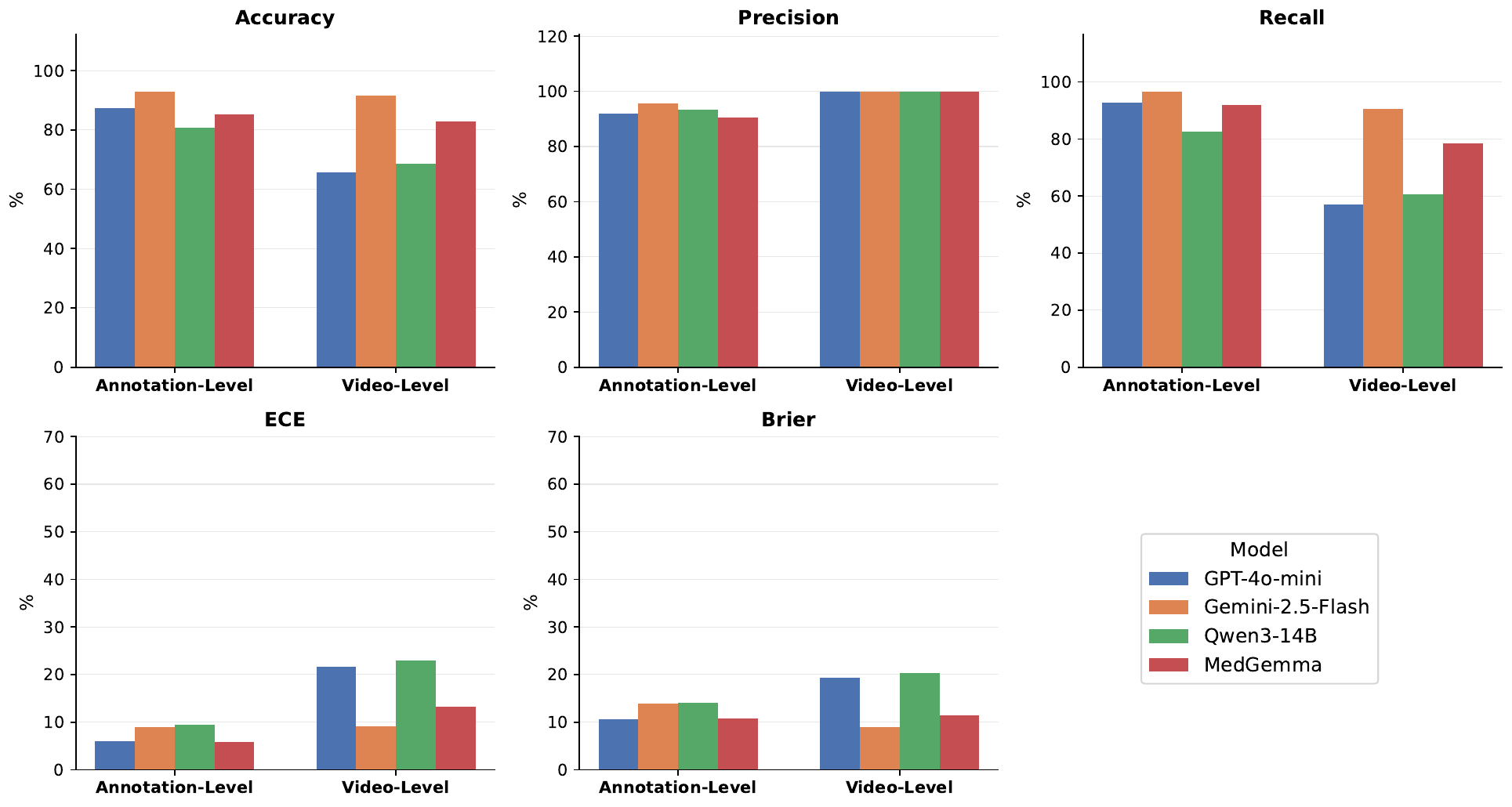}
    \label{fig:bar_plots_inacc}
    \caption{Figure \ref{fig:llm_eval_inacc} as bar plots}
\end{figure*}

\begin{figure*}
    \includegraphics[width=0.8\linewidth]{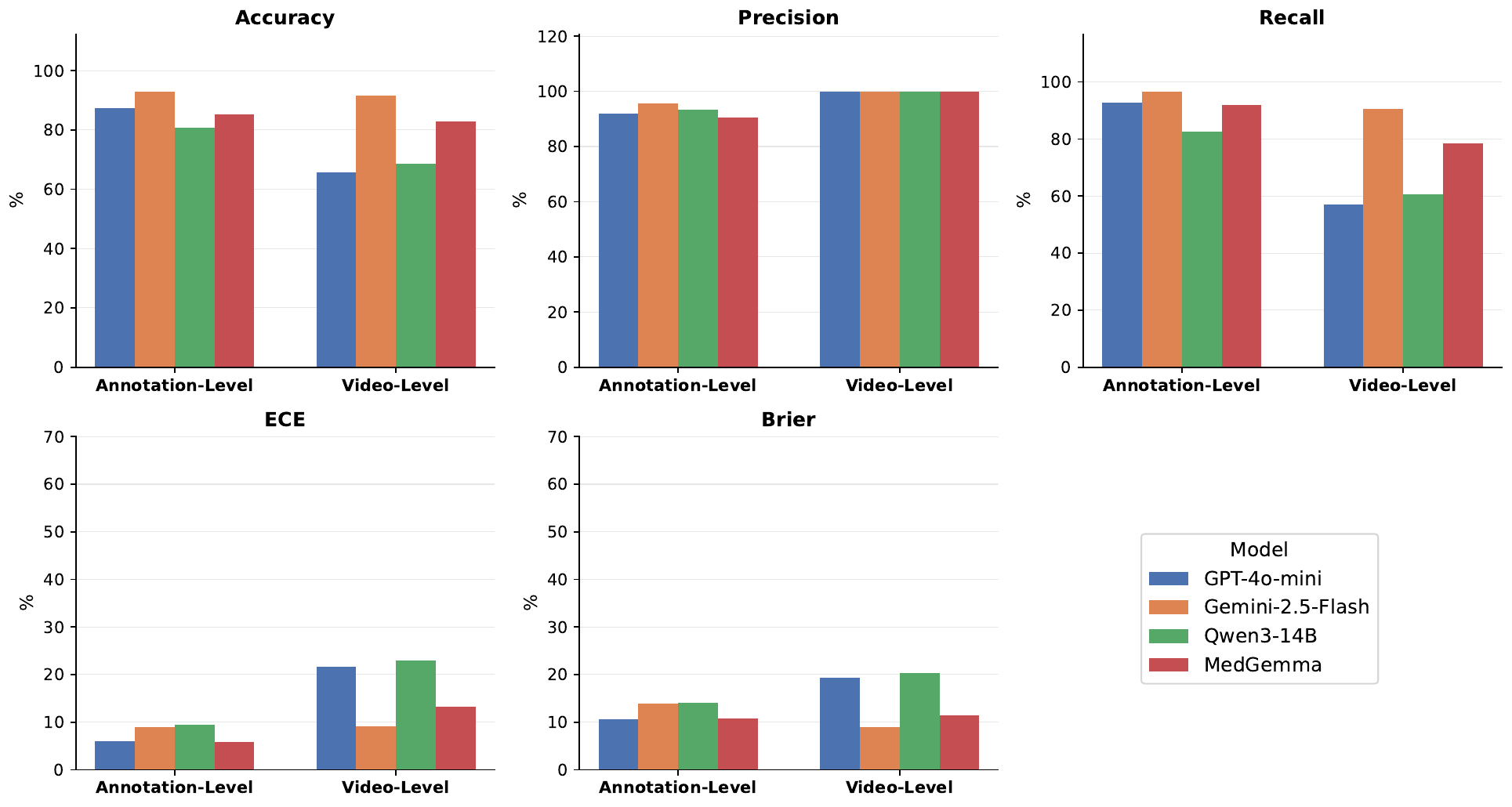}
    \label{fig:bar_plots_inacc}
    \caption{Figure \ref{fig:llm_eval_harm} as bar plots}
\end{figure*}

\section{Additional Artifacts and Analysis}
\label{sec:appendix_additional_analysis}

\begin{biggreenbox}[Top 50 Hashtags]
pregnancy, pregnant, obgyn, fyp, laboranddelivery, pregnancyjourney, pregnancytiktok, pregnancytips \\
birth, pregnanttiktok, pregnantlife, thirdtrimester, momsoftiktok, weekspregnant, postpartum, firsttimemom \\
baby, pregnancynutrition, doctorsoftiktok, prenatalnutrition, womenshealth, labor, firsttrimester, doula \\
foryou, doctor, braxtonhicks, birthtok, childbirtheducation, secondtrimester, homebirth, childbirth \\
midwife, epidural, greenscreen, pregnancytok, naturalbirth, ttc, hospitalbirth, prenatal \\
newmom, birthtips, learnontiktok, labourandbirth, momtobe, weeks, fertility, infertility \\
neuraltubedefect, pregnancydietitian 
\end{biggreenbox}

\subsection{Tracing n-grams in the LLM responses}

OLMoTrace \cite{liu2025olmotrace}, built using infini-gram \cite{liu2024infini}, is a system that allows tracing spans of LLM responses back to their full training data. While OLMoTrace is intended for users with access to the LLM and its training data, we use it to analyze websites containing similar n-grams. Note that these findings do not indicate a causal link between the websites and the n-grams, but are provided to understand a possible source for the facts (in the n-grams) stated in the LLM response. We use \gptfour responses and extract the website domain of the document containing an exact span/n-gram match with 10 or more words, where an occurrence of a domain is weighed ([0,1)) with the relevance score of the document. Some of the top occurring categories of websites are health information blogs (\texttt{pnmag.com},\texttt{healthline.com}), Wikipedia, news organizations, and businesses selling products. Surprisingly, exact matches from reputed sources like the Mayo Clinic and ACOG have little presence.

\subsection{Qualitative Analysis}
\begin{figure}
    \centering
    \includegraphics[width=0.8\linewidth]{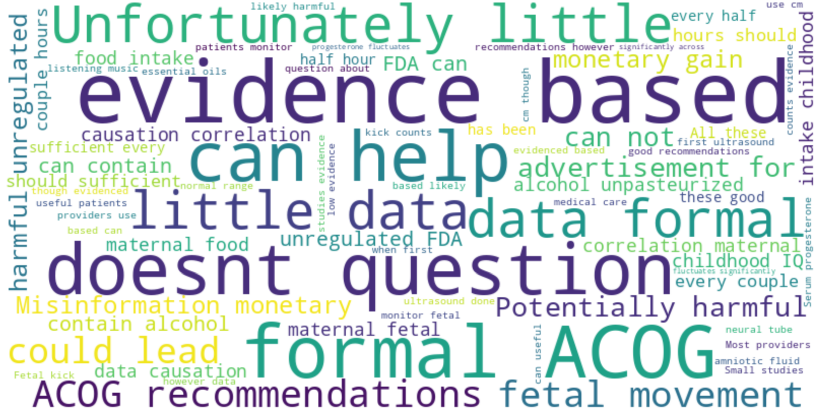}
    \caption{Word cloud of bigrams extracted from the reasons.}
    \label{fig:reason_wordcloud}
\end{figure}

Figure \ref{fig:reason_wordcloud} shows the common bi-grams in the reasons specified by the experts. Lack of evidence and ACOG-based recommendations appear to be the most common phrases.

\paragraph{Annotator Agreement and Reasoning } We calculate annotator agreement per video to overcome disagreement over sentence selection subjectivity. After a manual analysis on 30 randomly sampled instances of disagreements, we find that the disagreements primarily occur while interpreting the information and best practices. For instance, an interpretation disagreement is as follows: One expert reasons the image that displays \q{..Oligohydramnios.. diagnosed with < 5 cm} as correct while the other expert specifies it to be a 2cm x 2cm pocket of amniotic fluid. Similarly, a best practices example: two experts disagreed on the claim of hot baths being dangerous, where one expert claimed increased risk of miscarriage and other claiming it is okay.

\subsection{Video metadata analysis}

\paragraph{Engagement-accuracy correlation} Table \ref{tab:correlation} shows the correlation between the video engagement metrics (views, likes, and followers) and the accuracy. We find that there is no strong correlation, indicating that inaccurate or harmful information can be just as popular as accurate or harmless information.

\begin{table}
\centering
\begin{tabular}{@{}
>{\columncolor[HTML]{FFFDFA}}l 
>{\columncolor[HTML]{FFFDFA}}l 
>{\columncolor[HTML]{FFFDFA}}l @{}}
\toprule
{\color[HTML]{333333} \textbf{Engagement Metric}} & {\color[HTML]{333333} \textbf{Label}} & {\color[HTML]{333333} \textbf{Correlation (r)}} \\ \midrule
{\color[HTML]{333333} Views}     & {\color[HTML]{333333} Accuracy} & {\color[HTML]{333333} -0.043} \\
{\color[HTML]{333333} Views}     & {\color[HTML]{333333} Harm}     & {\color[HTML]{333333} 0.047}  \\
{\color[HTML]{333333} Likes}     & {\color[HTML]{333333} Accuracy} & {\color[HTML]{333333} -0.032} \\
{\color[HTML]{333333} Likes}     & {\color[HTML]{333333} Harm}     & {\color[HTML]{333333} 0.032}  \\
{\color[HTML]{333333} Followers} & {\color[HTML]{333333} Accuracy} & {\color[HTML]{333333} -0.025} \\
{\color[HTML]{333333} Followers} & {\color[HTML]{333333} Harm}     & {\color[HTML]{333333} 0.031}  \\ \bottomrule
\end{tabular}
\caption{Correlation between presence of accurate and harmful information with engagement metrics of a video.}
\label{tab:correlation}
\end{table}

\section{LLMs on various platforms}
\label{sec:appendix_llm_platforms}

\begin{figure*}
    \centering
    \begin{subfigure}[t]{0.23\textwidth}
        \centering
        \includegraphics[width=\linewidth]{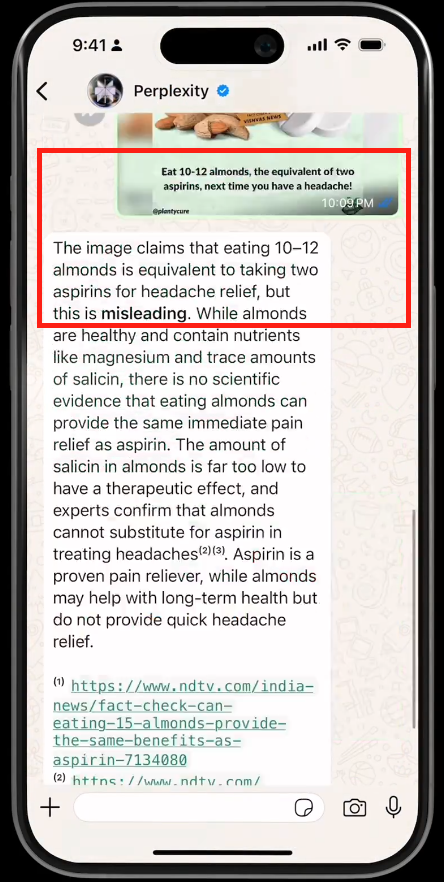}
        \caption{Perplexity on WhatsApp promoting fact-checking.}
        \label{fig:perplexity_llm}
    \end{subfigure}
    \hfill
    \begin{subfigure}[t]{0.23\textwidth}
        \centering
        \includegraphics[width=\linewidth]{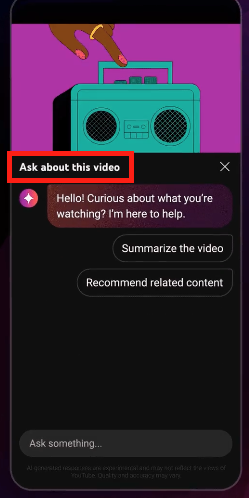}
        \caption{YouTube promoting users to ask questions related to the video.}
        \label{fig:youtube_llm}
    \end{subfigure}
    \hfill
    \begin{subfigure}[t]{0.23\textwidth}
        \centering
        \includegraphics[width=\linewidth]{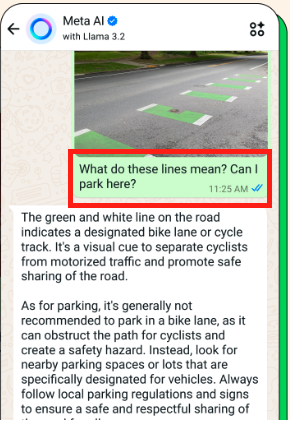}
        \caption{Meta AI on WhatsApp promoting users to seek answers about images.}
        \label{fig:whatsapp_llm}
    \end{subfigure}
    \hfill
    \begin{subfigure}[t]{0.23\textwidth}
        \centering
        \includegraphics[width=\linewidth]{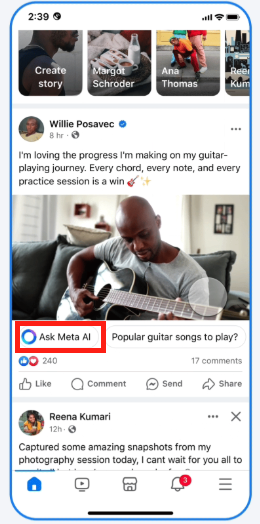}
        \caption{Meta AI on Facebook promoting users to ask about the post.}
        \label{fig:facebook_llm}
    \end{subfigure}
    \caption{Digital platforms promoting various kinds of information seeking behaviors \inlineredbox{(highlighted in a red box)}}.
    \label{fig:llm_digital_platforms}
\end{figure*}

In this section, we document the integrations of LLMs in various digital platforms. Figure \ref{fig:llm_digital_platforms} showcases how various platforms promote the possibility for users to engage in various types of information-seeking behavior. The sources

\begin{itemize}
    \item Perplexity on WhatsApp: \url{https://x.com/AskPerplexity/status/1917971703168401452}
    \item YouTube: \url{https://blog.youtube/inside-youtube/2024-in-youtube-ai/}
    \item Meta AI on WhatsApp: \url{https://blog.whatsapp.com/talk-to-meta-ai-on-whatsapp}
    \item Meta AI on Facebook:\url{https://about.fb.com/news/2024/04/meta-ai-assistant-built-with-llama-3/}
\end{itemize}

\section*{Broader Impacts}
\label{sec:broader_impacts}

Akin to most works in AI, our work has both positive and negative societal consequences, as it lies in the intersection of social media, health, and LLMs. Continually deploying our proposed methodologies can inspire future works to create dynamic datasets and benchmarks, as well as inform public health experts on current trends and focus areas. However, malicious actors can use similar strategies to design health disinformation that evades large language models. We strongly oppose any such actions and remain optimistic that parallel research in improving the security and reliability of LLMs can prevent such use cases.

\end{document}